\documentstyle[12pt,fleqn]{article}
\begin{document}
\sloppy

\begin{center}

{\Large{\bf GLOBAL ANISOTROPY OF SPACE AND EXPERIMENTAL INVESTIGATION OF
      CHANGES IN $\beta$-DECAY COUNT RATE OF RADIOACTIVE  ELEMENTS
}}
\vskip20pt

                          Yu.A. BAUROV

{\it     Central Research Institute of Machine Building,

141070, Korolyov,  Moscow Region, Russia}
\vskip10pt

                          A.A. KONRADOV

{\it Academy of Sciences, Institute of Biochemical Physics,

 117977, Moscow, Russia}
\vskip10pt

                 V.F. KUSHNIRUK and Yu.G. SOBOLEV

{\it Flerov Laboratory for Nuclear Reactions (FLNR),

Joint Institute for Nuclear Reactions,

141980, Dubna, Moscow Region, Russia}
\vskip25pt
 ABSTRACT
\end{center}
{\footnotesize
    The  results of experimental investigations of changes  in
  $\beta$-decay count rate of radioactive elements, are presented,
  and  an  explanation of those on the base of a new  physical
  conception  of forming the observed three-dimensional  space
  from  a  finite  set  of one-dimensional discrete  vectorial
  objects  (byuons),  containing  the  cosmological  vectorial
  potential  ${\bf A}_g$,  a  new  fundamental vectorial  constant,  is
  given.  In  the theory, the vector ${\bf A}_g$ direction  corresponds
  with  that  of the axis of Universe rotation being discussed
  in literature. }
\vskip10pt
 PACS numbers: 11.23
\vskip25pt
{\bf I. Introduction }
\vskip10pt
 Since  the  works  of  Birch [1,2], the  question  of  a  possible
rotation  of  the  Universe  is under discussion  in  scientific
publications  (even Aristoteles had touched on  this  subject [3]).
When  observing a correlation of polarization angles  for  light
from  distant sources, Birch anticipated that the Universe might
have an axis of rotation with coordinates $\alpha \approx 180^\circ\pm 30^\circ,
\delta \approx 35^\circ\pm 30^\circ$  in the second equatorial system.
In Ref. [4], these are presented, too ($\alpha \approx 315^\circ\pm 30^\circ,
\delta \approx 0^\circ\pm 20^\circ$), which differ, as is seen, from  those  of
Birch, and are doubted by the authors of this article because of
the  arrangement of the Universe's rotation axis in plane of the
celestial equator.

In  the  new  physical conception [5,6]  of forming  the  observed
three-dimensional space $R_3$ from a  finite set of one-dimensional
discrete  vectorial  objects  (byuons)  as  a  result  of  their
dynamics,  the  global  anisotropy of  the  Universe  originates
necessarily  and is caused by the direction of the  cosmological
vectorial  potential ${\bf A}_g$, a  new fundamental  vectorial  constant
appearing in definition of the byuons. Through minimization  of
the potential energy of byuon interaction in the one-dimensional
space  formed by them, rotation of the observed objects  and  of
the  Universe  as  a whole, comes into existence.  The  axis  of
rotation  of  the  Universe is to be  perpendicular  to  the  ${\bf A}_g$
vector which, according to experimental investigations with  the
use  of   high current magnets [7-9] and a gravimeter joint with  a
magnet [10],  as  well   as  in  agreement   with   astrophysical
researches [11,12], has coordinates  $\alpha \approx 270^\circ\pm 7^\circ,
\delta \approx 34^\circ$.  These  values are in good correspondence
with the results of Birch.

   According  to the theory [5], the value $ A_g \approx 1.95\times 10^{11}$
 CGSE  units is  the  limiting one. In reality, there exists, in the vicinity
of  the Earth, a certain summary vectorial potential ${\bf A}_\Sigma$, i.e. the
vectorial potentials of magnetic fields from the Earth ($A_E \le 10^8$
CGSE units), the Sun ($A_\odot \approx 10^8$ CGSE units), the Galaxy
($ \sim 10^{11}$ CGSE units),  and the Metagalaxy ($ > 10^{11}$ CGSE units)
are superimposing on the constant vector ${\bf A}_g$  resulting probably
in some turning of ${\bf A}_\Sigma$  relative to ${\bf A}_g$ in the $R_3$
  space, or in a decrease of ${\bf A}_\Sigma$.

Hence  in  the  theory  of physical space (vacuum),  which  the
present article leans upon, the field of the vectorial potential
introduced even by Maxwell, gains a fundamental character. As is
known,  this  field  was  believed as an  abstraction.  All  the
existing  theories  are  usually gauge  invariant,  i.e.
a  vectorial potential {\bf A} (  for example,  in classical and quantum electrodynamics)
is defined with an accuracy of an arbitrary function
gradient,  and  a scalar one is with that of time derivative  of
the  same  function, and one takes as real only  the  fields  of
derivatives of these potentials, i. e. magnetic flux density and
electric field strength.

 ln  Refs. [5,13] local violation of gauge invariance was supposed,
and  the elementary particle charge and quantum number formation
processes   were  investigated  in  some  set,   therefore   the
potentials  gained unambiguous physical meaning  there.  In  the
present paper, this set is a finite set of byuons.

The  works by D.Bohm and Ya.Aharonov [14], discussing the  special
meaning  of potentials in quantum mechanics, are the most  close
to  the  approach  under consideration, they  are  confirmed  by
numerous experiments [15].

 In  the  Refs. [13,16]  which are generalized by the byuon theory,
the probability $W$ of the $\beta$-decay of radioactive elements is shown
to  be  proportional to the summary potential $A_\Sigma$ into which  also
the   ${\bf A}_g$    vector  enters.  Hence  from  changes  in  $A_\Sigma$  and,
correspondingly,  in $W$, the direction of  ${\bf A}_g$  in  space  may  be
judged  despite  the fact that, as was thought earlier,  the  $\beta$-decay
 of  the  radioactive elements practically  could  not  be
effected  by  any physical influence (pressure, magnetic  field,
etc. [17]).
\vskip20pt

{\bf II. Experiments}
\vskip10pt

 By  now three runs of experiments have been performed to verify
this    assumption,   initially   at   Sankt-Petersburg    State
Technological University in 1994 (Fig.1) [18].

 The  magnitude of oscillations for $^{137}Cs$ equaled $\sim 25$  thousand
photons (with a starting level of $\sim 18.6$ million) in a period of
$\sim  16$  min., which corresponded to $\sim 6 \sigma$ (the standard  deviation
$\sigma = 4307$). The repetition of the effect throughout the whole cycle
of measurements proved  it was not accidental.
The  minimum intensity  of $\beta$-decay corresponded  to $\sim 10^h$  of
Moscow time ($8^h$ of astronomical time, i.e. $8^h$ a.m.).

 Concurrently,   similar  measurements  were  conducted   during
several  days  with  the  use  of $^{60}Co$  (see  Fig.1).  Analogous
oscillations  had an amplitude of $20$ thousand events  (with  the
average level of $8.6\times10^6$) in an exposure time of $\sim  7$ min,  i.e.
approximately $7\sigma$ .

In  August-September  1996 and in February  1997  two  runs  of
experiments  on  investigation of $^{90}Sr$  $\beta$-decay  count  rate  were
carried  out  at  FLNR  JINR (Dubna) (1st and  2nd  experiments,
respectively).

The  electrons  were  detected by a fast scintillation
counter (with a crystal YAG:Ce) optically coupled to a low noise
photomultiplier FEU-143PM with a standard divider.

In each measurement the following information was recorded:

 -  the number of pulses counted by the electron detector  in  1s;

-  the number of pulses arrived from the "Quartz $10^6$" generator
and necessary to control the exposure time;

 -  the  astronomical  time  of   the  measurement  of  interest
obtained from a timer.

 The  entire  set  was  situated in a  thermostable  room  where
fluctuations in temperature were no more than $2^\circ C$ for  the  whole
time of an experiment. Before a session of measurements, the set
was  forced into an operating regime in a week. The high voltage
applied to the photomultiplier from a HV unit (= 1kV) of POLON 1904 type
as  well as voltages of powering CAMAC units in a crate, checked
periodically  during the entire measurements, were  stable,  the
first one with an accuracy of   1V, the next ones ($\pm 24$ and $\pm 6$
 V) with that of $\pm1\%$.
\vskip20pt

{\bf III. Analysis of experiments}
\vskip10pt

 The  analysis  was  made with the purpose  of  determining  the
presence of a daily periodicity and a degree of irregularity  in
distribution  of  decay  number over the astronomical  day.  The
initial  sequence of every second measurements had a  length  of
nearly $1.2\times 10^6$ points, and formed a base for next computations.
Since  we  were  interested  in almost  daily  periodicity,  the
initial sequence was averaged over the minute period. The result
is  shown in Fig.2. Seen is a slow dynamics that may be  related
to  the  presence  of long-period oscillations.  Those  of  high
frequency  were viewed, too, but the daily rhythm could  not  be
perceived in such a form at once. The dynamics as a whole is  of
complicated  character, therefore to analyze a daily  component,
the  slow  (two-weekly)  nonlinear trend  was  excluded  through
selecting  the best function from the large family of those  and
subtracting  it from the series (2). The presence of  periodical
components in the remainder (after withdrawal of the trend,  see
Fig. 3) was revealed by two methods.

 First,  the  standard Fourier analysis of the  series  (3)  was
carried out. In Fig. 4 where the results are presented, at least
two  frequencies stand out, of which the former  ("great  peak")
corresponds  to approximately half-week period, and  the  latter
("small peak") does to the daily one. The half-week period  will
be  considered  subsequently. The  daily  periodicity  found  by
Fourier-method   gives,  however,  no  way  to   determine   the
astronomical time of an event when the measured value is greater
or less than the average.

 Second,  the  following  procedure  was  used  to  refine   the
distribution of $\beta$-decay  numbers over the astronomical day.  Each
moment (minute) of measurement was represented as a point  of  a
circle  (corresponding to an astronomical day) and expressed  in
degrees, so that the whole series could be "coiled" around  that
circle. Thus, each measurement was related to a certain time  of
day (in degrees). If the quantity to be measured was uniform  in
time,  then  the distribution thereof over the circle  would  be
uniform,  and the hypothesis for uniformity of said distribution
could  be  verified by statistical methods. For this purpose  we
used  the  Kolmogoroff-Smirnoff's test based on  computation  of
maximum  difference  between  the theoretical  and  experimental
distribution  functions and on comparing  it  with  a  tabulated
value.  If  this difference was sufficiently large for  a  given
sample  size and an accepted significance level, the  conclusion
was  drawn  that the experimental and theoretical  distributions
were  different. In our case, with the theoretical  distribution
chosen being uniform, this signifies the nonuniform distribution
of the metered quantity in time. The Kolmogoroff-Smirnoff's test
is  convenient  also  in  view of obtaining,  in  parallel  with
estimation of confidence level,  a point (day time) at which the
deviation from uniformity is maximum.

According to our conception of influence of changes in $A_\Sigma$   on
the $\beta$-decay  count rate, we are interested primarily in values of
fluxes  lesser  than the average indication  of  the  instrument
because  the modulus $|{\bf A}_\Sigma|$  is always below  $|{\bf A}_g|$.
 On this  basis,
when  analyzing  a distribution, we take into account  only  the
values  lesser  than  the average by $L\times\sigma$  where  $L$
 is  a  factor
determining the extent of deviation from the average.  For  each
such an extreme value, one notes the point in time at which that
takes  place,  and  tests  the  hypothesis  for  uniformity   of
distribution of those points over the astronomical day expressed
in degrees (24 hours are equal to $360^\circ$).

In   Fig.5    are  graphically  represented  the   results   of
computation  for $ L = 2$. The $X$ axis is  astronomical  time  in
degrees  (from  0  to  $360^\circ$), the $Y$ axis is  deviation  from  the
uniform  distribution.  The confidence interval  for  $P=0.05$  is
shown dashed. It is clearly seen that the frequency function  of
the  sample is highly nonuniform and peaks at an angle of  about
$90^\circ$. At this point, the $5\%$ level of significance is far exceeded.

Since  the  counting  in the experiment  on  23rd  August  1996
began at $20^h$ of Moscow time (i.e. at $18^h$ of astronomical time),
the  angle  of $90^\circ$ corresponds to $24^h$  of astronomical time
 ($12^h$ p.m.). In Fig.6 the indicated point of time is asterisked.

In  the  experiment carried out in February 1997, in connection
with  the enormous massive of data obtained in the previous one,
the events were recorded not at secondly intervals but every  10
seconds.  The results of this experiment, processed analogously,
are shown in Fig.7. The experiment began on 22nd February at $5^h 21^m$
 of Moscow time. In Fig.8 is shown the Fourier spectrum of
the  same signal as in Fig.7. As in the preceding experiment,  a
huge  peak corresponding to half-week period as well as  a  less
significant  one corresponding to 24-hour period, stand  out.  A
significant difference was the fact that in this experiment  the
events  beyond the scope of $2\sigma$ in direction of decay  moderation
practically  did not observed, but did those beyond $2\sigma$  directed
to  increase  in $\beta$-decay  rate. In Fig. 9, the maximum  of  said
events  corresponds to an angle of $280^\circ$, i.e. to a shift in  time
of about $19^h$ from the start ($23^h 21^m$ of astronomical time, i.e.
$11^h 21^m$ p.m.). In Fig. 6, the point in time corresponding to  the
maximum  intensity  of $\beta$-decay within the  24-hour  observation
period,  is  asterisked. By (*) denoted are in this  Figure  the
times  of  maximum deviations from the average value of  $\beta$-decay
intensity  of  radioactive  elements  for  all  three  runs   of
experiments,  such  points  are coincident  with  the  times  of
maximum changes of $A_\Sigma$ due to the vectorial potential of the Earth
${\bf A}_E$.  However,  in  the experiment of February,  we  observed  an
increase in $\beta$-decay count rate at the point of extremum, but not
the decrease, as in August-September 1996, which
may  be  explained  by  the fact that in  February $A_\Sigma$  had  been
augmented  due to a decrease in the vectorial potential  of  the
Sun ${\bf A}_\odot$  by  that of the Earth ${\bf A}_E$, which is clearly seen  in  the
Fig.6.

For the ultimate conclusions, the experiments of 1996-97 on investigating the
time changes in the decay rate of radioactive elements were replicated in
March 1998. In Fig.10 given are the temporal series of $^{60}Co$ $\beta$-decay
intensity values, smoothed out by the moving average with a window of $\sim 100s$,
as well as the results of evaluation of irregularity obtained by the same procedure
as when analyzing experiments of August-September 1996 and February 1997.
The Fourier-analysis has shown presence of a clearly defined near-daily period
with a phase of maximum decrease in $\beta$-decay intensity (from the beginning of
the experiment) being in the region of around $200\div 230$ degrees (Fig.11).
In Fig.6 the said maximum is asterisked, this point corresponds to the maximum
decrease in ${\bf A}_\Sigma$ due to the vectorial potential ${\bf A}_E$.

 Thus,  the analysis of 24-hour variations in the count rate  of
$\beta$-decay of radioactive elements for all four runs of experiments
has confirmed our theoretical assumptions and corresponds to the
direction of ${\bf A}_g$ vector experimentally determined earlier [7-9].

Fig.12  demonstrates  the results of  measuring  noise  in  the
experimental  set,  which  were obtained  through  a  two-weekly
observation of the background at the experiment place without  a
radioactive  source. As is seen in Fig.12,  a  weekly  drift  in
readings  similar  that  of Fig.2, takes place.  Therefore,  the
weekly  drop in the decay intensity observed in August-September
1996,  was probably caused exactly by the drift in the set.  The
background  may  not act on the position of the extremum  points
(*)  in  Fig.6,  because the flux $Y$ from that is two  orders  of
magnitude lesser than from the radioactive source. Therewith the
number of decay events beyond $2\sigma$ is nearly an order of magnitude
greater than the similar one in the background (Fig.12).

 Now  let  us discuss the half-weekly peak obtained on  base  of
Fourier analysis of signals in the experiments. To reach a  sure
conclusion  about  reasons  of its origin,  an  experiment  with
duration   of  at  least  a  month  is  of  a  prime  necessity.
Nevertheless,  proceeding from the sectorial  structure  of  the
interplanetary  magnetic  field (one  sector  corresponds  to  a
week), we may assume that the 3.5-daily period of  $\beta$-decay  count
rate  variation, found by us, is probably related  just  to  the
behavior  of the Sun's vectorial potential. In all  likelihood,
we  are dealing here not with that related to the interplanetary
magnetic fields blown by the solar wind plasma fluxes, but  with
a  vectorial potential originated within the Sun and  distorting
the  physical space around it, i.e. changing ${\bf A}_\Sigma$ entering into the
definition of  byuons.

 Thus,  the  experiments  carried out have  shown  the  relation
between  the global anisotropy of space caused by the  existence
of   ${\bf A}_g$  and  the  count  rate  variations  in  the  $\beta$-decay
 of radioactive elements.
\vskip20pt
{\bf Acknowledgments}
\vskip10pt
 The    authors   would   like   to   thank   acad. S.T. Beljaev,
Yu.Ts. Oganesjan, Yu.E. Penionzhkevich, V.V. Dvoeglazov
for supporting their work and valuable discussions.

\vskip10pt

{\bf Subscripts to drawing}
\vskip10pt

 Fig.1.  The changes of $^{137}Cs$, $^{60}Co$ $\beta$-decay count rate
in  time in experiments 19-23.04.94;  a -  137Cs;    b  -  60Co

 Fig.2.   The  initial  sequence  averaged  over  the   minute's
periods;    t -  time in weeks (1st experiment)

 Fig.3.  The  normalized  count rate averaged  over  1  minute
     versus  time after extracting of the nonlinear  trend  from
     the initial series of measurements (1st experiment).

 Fig.4.   The Fourier spectrum of the signal shown in Fig.3 (1st
     experiment).  The  first  peak  corresponds  to   half-week
     period, the second one corresponds to 24-hour period.

 Fig.5. The dynamics of the ununiformity for
            a  24-hour period (1st experiment).
           $Y$  is the Kolmogorov statistic value.

 Fig.6.   The   directions  of  the  ${\bf A}_g$  vector  and   vectorial
     potentials   of   magnetic  fields   from   the   Sun  ${\bf A}_\odot$
     and the Earth ${\bf A}_E$.

$8^h$, (*) - astronomical time points corresponding to
extremum changes in $\beta$-decay count rate.

Fig.7. The same as in Fig.3  for the 2nd experiment.

Fig.8. The same as in Fig.4  for the 2nd experiment.

Fig.9. The same as in Fig.5  for the 2nd experiment.

Fig.10. The experiment on 15-19 March 1998. Start in $1^h$ of Moscow time
($0^h$ of astronomical time, i.e. $0^h$ a.m.). The temporal series of $^{60}Co$
$\beta$-decay intensity values, smoothed out by the moving average with a window
of 100s. $N_1$ - moving average of intensity, $t$ - time in 100s.

Fig.11. Daytime dependence of the relative frequence of events $F(\theta)$
going out of $2\sigma$ ($\theta$ in degrees).

Fig.12.  The  results  of measuring noise in  the  experimental set.
\end{document}